\documentstyle[11pt,aaspp4]{article}

\begin{document}

\renewcommand{\thefootnote}{\fnsymbol{footnote}}

\title{Radial Velocity Studies of Close Binary 
Stars.~III\footnote[1]{Based on the data obtained at the David Dunlap 
Observatory, University of Toronto.}}

\renewcommand{\thefootnote}{\arabic{footnote}}

\author{Slavek M. Rucinski, Wenxian Lu, Stefan W. Mochnacki\\
e-mail: {\it rucinski@astro.utoronto.ca, lu@astro.utoronto.ca,
mochnacki@astro.utoronto.ca}}

\affil{David Dunlap Observatory, University of Toronto \\
P.O.Box 360, Richmond Hill, Ontario L4C~4Y6, Canada}

\centerline{\today}

\begin{abstract}
Radial velocity measurements and simple sine-curve fits to the orbital 
velocity variations are presented for the third set of ten contact binary 
systems: CN~And, HV~Aqr, AO~Cam, YY~CrB, FU~Dra, RZ~Dra, UX~Eri, 
RT~LMi, V753~Mon, OU~Ser. All systems but two are contact, double-line 
spectroscopic binaries with four of them 
(YY~CrB, FU~Dra, V753~Mon, OU~Ser) being the recent
discoveries of the Hipparcos satellite project. 
The most interesting object
is V753~Mon with the mass-ratio closest to unity among all contact
systems ($q = 0.970 \pm 0.003$) and large total mass 
($(M_1+M_2) sin^3i = 2.93 \pm 0.06$). Several of the studied 
systems are prime candidates for combined light and 
radial-velocity synthesis solutions.
\end{abstract}


\section{INTRODUCTION}
\label{sec1}

This paper is a continuation of the radial velocity studies of close 
binary stars, \cite{LR99} (Paper I) and \cite{RL99} (Paper II).
The main goals and motivations are described in these papers.
In short, we attempt to obtain modern radial velocity data for
close binary systems which are accessible to the 1.8 meter class 
telescopes at medium spectral resolution of about R = 10,000 -- 15,000. 
Selection of the objects is quasi-random in the sense that we
started with shortest period contact binaries. The intention is
to publish the results in groups of ten systems as soon as reasonable
orbital elements can be obtained from measurements evenly distributed 
in orbital phases. We are currently observing a few dozens of such systems.
The rate of progress may slow down, as we move into systems with
progressively longer periods. 
  
This paper is structured in the same way as Papers~I and II in that it 
consists of two tables containing the radial velocity 
measurements (Table~\ref{tab1}) and 
their sine-curve solutions (Table~\ref{tab2}) 
and of brief summaries of previous studies for individual systems. 
The reader is referred to the previous papers for technical details 
of the program. In short, all observations described here were 
made with the 1.88 meter telescope of the David 
Dunlap Observatory (DDO) of the University of Toronto. The 
Cassegrain spectrograph giving the scale of 
0.2 \AA/pixel, or about 12 km/s/pixel, was used; 
the pixel size of the CCD was 19~$\mu$m. A relatively wide 
spectrograph slit of 300~$\mu$m corresponded to the angular size on 
the sky of 1.8 arcsec and the projected width of 4 pixels.   
The spectra were centered at 5185 \AA\ with the spectral coverage of 
210 \AA. The exposure times were typically 10 -- 15 minutes long.

The data in Table~\ref{tab1} are organized in the same manner as 
in Paper~II. It provides information about the relation between 
the spectroscopically observed epoch of the primary-eclipse T$_0$ 
and the recent photometric determinations in the form of the (O--C) 
deviations for the number of elapsed periods E. It also contains, 
in the first column below the star name,
our new spectral classifications of the program objects. The classification
spectra, were obtained with a grating giving a dispersion of 
0.62 \AA/pixel in the range 3850 -- 4450 \AA. The program-star spectra 
were ``interpolated'' between spectra of standard stars 
in terms of relative strengths of lines 
known as reliable classification criteria.

In the radial-velocity solutions of the orbits, the data have  
been assigned weights on the basis of  our ability to resolve the 
components and to fit independent Gaussians to each of the 
broadening-function peaks. Weight equal to zero in 
Table~\ref{tab1} means that an observation was not used 
in our orbital solutions; however, these observations may be utilized in 
detailed modeling of  broadening functions, if such are 
undertaken for the available material. The full-weight points are 
marked in the figures by filled symbols while 
half-weight points are marked 
by open symbols. Phases of  the observations with zero weights are 
shown by short  markers in the lower parts of the figures; they were 
usually obtained close to the phases of orbital conjunctions.

All systems discussed in this paper but one have been observed for
radial velocities for the first time. The only exception is RZ~Dra
for which an SB1 orbit solution was obtained by \cite{str46}.
The solutions presented in Table~\ref{tab2} for 
the four circular-orbit parameters, $\gamma$, K$_1$, K$_2$ and T$_0$, 
have been obtained iteratively, with fixed values of the orbital period. 
First, two independent least-squares solutions for 
each star were made using the same programs as described in Papers~I
amd II. Then, one combined solution for both amplitudes and the common 
$\gamma$ was made with the fixed mean 
value of T$_0$. Next, differential corrections for $\gamma$, 
K$_1$, K$_2$, and T$_0$ were determined, providing best 
values of the four parameters. These values are given in Table~\ref{tab2}. 
The corrections to $\gamma$, K$_1$, K$_2$, and T$_0$ were finally 
subject to a ``bootstrap'' process (several thousand 
solutions with randomly drawn data with repetitions) to 
provide the median values and ranges of the
parameters. We have adopted them as measures of  
uncertainty of parameters in Table~\ref{tab2}.

Throughout the paper, when the errors are not written otherwise, we 
express standard mean errors in terms of the last quoted digits, 
e.g. the number 0.349(29) should be interpreted as $0.349 \pm 0.029$. 

\placetable{tab1}

\placetable{tab2}

\section{RESULTS FOR INDIVIDUAL SYSTEMS}
\label{sec2}

\subsection{CN And}

The variability of CN~And was discovered by \cite{hof49}. Modern 
light curves were presented by \cite{kal83}, \cite{raf85},
\cite{kes89} and \cite{sam98}. The light curve is of the EB-type,
with unequally deep eclipses. From the eclipse timing, 
there exists a clear indication of a
continuous period decrease. We used the almost contemporary
with our observations determination of the primary eclipse time 
$T_0$ by \cite{sam98}. The radial velocity curve
of the secondary component (see Figure~\ref{fig1})
shows some asymmetry in the first half of the orbital period which 
may explain why our $T_0$ is shifted by $-0.014$ day to earlier
phases. From the radial-velocity point of view, the system
looks like a typical A-type contact binary, but it can be also
a very close semi-detached system.

Two groups of investigators attempted solutions of the light curves
for orbital parameters, disregarding lack of any information on
the spectroscopic mass-ratio. Our $q_{sp} = 0.39 \pm 0.03$ differs
rather drastically from these photometric estimates.
\cite{kal83} expected the mass-ratio to be within 
$0.55 < q_{ph} < 0.85$. \cite{raf85} found the most 
likely interval to be $0.5 < q_{ph} < 0.8$; 
they saw indications of a strong contact. The matter
of the contact or semi-detached nature of the system
should be re-visited in view of the 
discrepancy between our $q_{sp}$ and the previous
estimates of $q_{ph}$. 

The photometric data at maximum light were published by \cite{kal83}:
$V = 9.62$, $(B-V) = 0.45$. The system is apparently a moderately
strong X-ray source (\cite{sha96}). 

\placefigure{fig1}

\subsection{HV Aqr}

Variability of HV~Aqr was discovered relatively recently 
by \cite{hut92}. The type of variability was identified by
\cite{sch92} and a preliminary, but thorough study was
presented by \cite{rob92}. The system shows total
eclipses at the shallower of the two minima, so it is definitely
an A-type one. This circumstance is important to our
results because both available 
ephemerides of Schirmer and Robb do
not predict the primary minimum correctly and we are
not sure how our radial-velocity 
observations relate to the photometric observations: Our $T_0$
falls at phase 0.47 for the former and at 0.70
for the latter. Probably the values of the orbital period
are slightly incorrect for both ephemerides. We used
the period of 0.374460 day, following \cite{sch92}; if
this is incorrect, then part of the scatter in our data
may be due to the incorrect phasing over the span of 450 days
of our observations. 

The mass-ratio of HV~Aqr is small, $q_{sp} = 0.145 \pm 0.05$.
It agrees perfectly with the photometric solution of
\cite{rob92} of $q_{ph} = 0.146$ which confirms the validity
of the photometric approach for totally eclipsing systems,
in contrast with the typical lack of agreement
for partially eclipsing systems.

Our spectral type of F5V agrees with the observed $(B-V)=0.63-0.78$
for a relatively large reddening of $E_{B-V}=0.08$
expected by \cite{rob92}. The system is bright with $V = 9.80$,
and is potentially one of the 
best for a combined spectroscopic -- photometric solution.

\subsection{AO Cam}

Variability of AO Cam was discovered by \cite{hof66}. \cite{mil82}
analyzed their photometric observations in a simplified way.
Subsequent photometric solution by \cite{eva85}
and even the sophisticated work of \cite{bar93} 
did not bring much progress in view of the partial eclipses
and total lack of any information on the mass-ratio. \cite{bar93}
stated that AO~Cam is definitely a W-type system; they 
estimated the photometric mass-ratio at $q_{ph}
=1.71 \pm 0.04$ or $1/q = 0.585$. This value is very different
from our spectroscopic result, $q_{sp} = 0.413 \pm 0.011$,
but we do confirm the W-type of the system.

AO Cam does not have any $UBV$ data. Our spectral
classification is G0V. In view of its
considerable brightness of $V = 9.50$ at maxima, 
it appears to be a somewhat neglected system. 

To predict the moment of the primary eclipse $T_0$,
we used the ephemeride based on the period of \cite{eva85}
and the observations of the {\it secondary eclipses\/}
by \cite{fau86}. The (O--C) deviation is relatively small
in spite of many orbital periods elapsed since Faulker's
observations.

\subsection{YY CrB}

YY~CrB is one of the variable stars discovered by the Hipparcos
satellite mission (\cite{hipp}). The only available 
light curve comes from this satellite; 
the star has not been studied in any other way.
There are no obvious indications of total eclipses, but
the coverage of eclipses is relatively poor so that
it is possible that the minimum identified as the primary
is not the deeper one. We used the primary minimum ephemeride of
Hipparcos and this results in a contact system of type A, that is with
the more massive and hotter component eclipsed at this minimum.

The system is bright, $V_{max} = 8.64$, and its Hipparcos parallax is
relatively large and well-determined, $p = 11.36 \pm 0.85$
milli-arcsec (mas), giving a good estimate of the
absolute magnitude of the system, $M_V=3.92 \pm 0.16$. With
$(B-V) = 0.62 \pm 0.02$ from the Hipparcos database, the
$M_V(\log P, B-V)$ calibration (\cite{RD97}) gives 
$M_V^{cal}=3.88$ so that the agreement is perfect. Our 
spectral classification of F8V agrees with the $(B-V)$ color index.

With the very good parallax and the new radial velocity data,
YY CrB is one of the systems which have a great 
potential of providing an excellent 
combined photometric and spectroscopic solution.

\subsection{FU Dra}

FU~Dra is another Hipparcos discovery. Again, we assumed the
primary eclipse identification and its ephemeride as in the 
Hipparcos publication (\cite{hipp}). With these assumptions, the system
appears to belong to the W-type systems. The mass-ratio
is somewhat small for such systems, $q_{sp} = 0.25 \pm 0.03$.

The system is bright, $V_{max} = 10.55$, but its Hipparcos parallax
is only moderately well determined, $p = 6.25 \pm 1.09$ 
mas, resulting in the absolute magnitude $M_V=4.53 \pm 0.38$.
The system was measured by the Hipparcos project
to have a relatively large tangential motion,
$\mu_{RA} = -255.85 \pm 1.18$ mas and $\mu_{dec} = 16.61 \pm 1.18$ mas.
The large proper motion had been noticed before (\cite{lee84a}, 
\cite{lee84b}), but the spatial velocity was then estimated 
assuming an uncertain spectroscopic parallax. Using the Hipparcos
data one obtains the two tangential components, $V_{RA} = -194$ 
km~s$^{-1}$ and $V_{dec} = 13$ km~s$^{-1}$. The radial
velocity $\gamma = -11$ km~s$^{-1}$ is moderate, so that only the
RA component of the spatial tangential velocity is very large. 

\placefigure{fig2}

\subsection{RZ Dra}

RZ Dra has been frequently photometrically observed 
since the discovery by \cite{cer07}.
The most recent extensive analysis of the system was by \cite{kre94}.
It utilized the only extant set of
spectroscopic observations by \cite{str46}
which led to a detection of one, brighter component. 
Our data confirm the primary amplitude $K_1 = 100$ km~s$^{-1}$, 
but we have been to also detect the secondary
component. The system consists of components considerably
differing in the effective temperature, and thus is classified as
an EB-type binary. Spectroscopically, one sees a more massive
component eclipsed at the deeper minimum so that it can be an
Algol semi-detached binary or an A-type contact system in poor
thermal contact.
The analysis of \cite{kre94} was made under the assumption 
of the semi-detached configuration. They found the 
photometric mass-ratio
$q_{ph} \simeq 0.45$, but some solutions suggested $q_{ph}
\simeq 0.55$. Our relatively well defined solution for both
components gives $q_{sp} = 0.40 \pm 0.04$. The same
investigation of Kreiner et al.\ 
provided the starting value of $T_0$. In spite of
indications that the period may be variable and that
many epochs elapsed since the study by Kreiner et al.,
the observed shift in the primary eclipse time is relatively 
small.

The spectral type that we observed, A6V, most probably applies to the
primary component which is much hotter than its companion (we
have not attempted to separate the spectra in terms of spectral types).
The Hipparcos parallax of the system, $p=1.81 \pm 1.01$ mas,
is too poor for an more extensive analysis of the absolute
magnitude of the system. RZ~Dra appears to
be a relatively short-period (0.55 day)
semi-detached Algol with both components accessible to
spectroscopic observations.

\subsection{UX Eri}

UX Eri is a contact binary which was extensively 
photometrically observed since its discovery by \cite{sol37}.
The first modern contact-model solutions which gave surprisingly
good agreement with our spectroscopic mass-ratio, $q_{sp}=0.37 \pm 0.02$,
was presented by \cite{mau72} almost quarter of a century ago;
it utilized the light curve of \cite{bin67} and
arrived at $q_{ph} = 0.42$. 

Our observations have been supplemented by four
observations obtained at the same time by Dr.\ Hilmar
Duerbeck at the European Southern Observatory with
a 1.52m telescope and a Cassegrain spectrograph. 
As a starting point of our solution, 
we used the moment of the primary minimum $T_0$ 
as predicted on the basis of observations 
by \cite{AH98a} and \cite{AH98b}; both took place actually
slightly after our spectroscopic observations.

UX Eri appears to be an A-type contact binary. 
The $(B-V)$ color index is not available for this star
(it has also not been measured by Hipparcos),
so we have not been able to relate it to our spectral 
classification of F9V.
The Hipparcos parallax $p=6.57 \pm 2.84$ mas provides
a relatively poor estimate of the absolute magnitude 
for the maximum brightness of $V_{max} = 10.59$:
$M_V = 4.7 \pm 0.9$.

\subsection{RT LMi}

RT LMi was discovered as a variable star by \cite{hof49}. The most recent
analysis from which we took the time of the primary eclipse $T_0$
is by \cite{nia94}. The system has been characterized in this study as 
a W-type contact binary of spectral type G0V with photospheric 
spots. Our spectral classification is based on poor spectra,
but they indicate a slightly earlier spectral type, of approximately
F7V. As \cite{nia94}
pointed out, the minima were observed to be of almost equal depth.
Our spectroscopic orbit gives an A-type so that the minimum selected
by the authors as the primary corresponds to the eclipse of the
more massive component (the temporal shift is very small
in spite of many elapsed epochs). The photometric solution of \cite{nia94}
assuming the W-type appears therefore to be invalid. 
The system is otherwise quite an inconspicuous
contact binary. It lacks even most essential photometric data.
The Simbad database gives $V_{max} = 11.4$, but the source of this
value is not cited.

\placefigure{fig3}

\subsection{V753 Mon}

V753 Mon is a new discovery of the Hipparcos mission. It is probably
one of the most interesting new close binaries recently discovered.
V753 Mon has not been studied before. The only published 
photometric data come from the $uvby$ survey 
of \cite{ols94} who found $(b-y) = 0.214 \pm 0.008$,
$m_1=0.160 \pm 0.003$ and $c_1=0.693 \pm 0.023$. The 
$(b-y)$ color index color corresponds
to $(B-V) \simeq 0.34$ or the spectral type F2V. Our spectral
classification is A8V which does not agree with these estimates
and with $(B-V)=0.36$ in
the Hipparcos catalog, unless there is considerable reddening of
about $E_{B-V} \simeq 0.12$. However, the early type would be
in a better accord with the large masses indicated by the
radial velocity solution (see below).
The brightness data in the Olsen measurements
indicated quite appreciable variability, $V=8.46 \pm 0.34$, but this
indication has been apparently overlooked. The Hipparcos mission
database treats it as a new discovery.

The two features distinguish V753 Mon as a particularly interesting
system: The mass-ratio close to unity and the large amplitudes of radial
velocity variations indicating a large total mass. The mass-ratio
$q_{sp}=0.970 \pm 0.009$ is the closest to unity of all 
known contact binaries. Note that the currently largest
mass ratios are $q=0.80$ for VZ~Psc (\cite{hri95})
and SW~Lac (\cite{zha89}) and $q=0.84$ for OO~Aql
(\cite{hri89}).
Since contact systems with $q \simeq 1$ are not observed, but 
are expected to experience {\it strong favorable 
observational biases for their detection and ease in analysis\/},
it is generally thought that contact configurations avoid this
particular mass-ratio.
The summed amplitudes of the radial velocity variations 
for V753~Mon give the
total mass of the system, $(M_1+M_2) sin^3i = 2.93 \pm 0.06$. 
This is in perfect agreement with the expected masses of two
main-sequence stars of the spectral type F2V seen on an orbit
exactly perpendicular to the plane of the sky. The light
variation is about 0.52 mag., in place of the expected
about 1.0 mag.\ for a contact system with $q \simeq 1$; 
therefore, the total mass for $i<90^\circ$ may turn 
out to be substantially larger. For the spectral type
estimated by us the individual masses should be close 
to $1.7\,M_\odot$.

Our radial velocity data show that
with the ephemeride based on the Hipparcos data, the system belongs
to the W-type contact systems. Apparently, the eclipses are
of almost the same depth, as expected for $q \simeq 1$.
However, the light curve from Hipparcos
is poorly covered around the secondary minimum so that 
the identification of the eclipses is uncertain. Obviously,
the distinction between A-type and W-type systems becomes
immaterial for $q \rightarrow 1$. 

The system is begging a new light curve and an extensive analysis,
not only because of the unusual properties, but also because it
is bright, $V_{max}=8.34$, and has a moderately well determined
Hipparcos parallax: $p=5.23 \pm 1.04$ resulting in
$M_V = 1.93 \pm 0.43$. This is in perfect agreement with
the $M_V(\log P, B-V)$ calibration which gives $M_V^{cal} = 1.90$
for the assumed $(B-V)=0.34$. The agreement would not be that good
if the color index is smaller, say 0.22, then one would obtain
$M_V^{cal} = 1.54$ which is still within the uncertainty 
of the parallax. Further investigations of V753~Mon will therefore
contribute to the absolute-magnitude calibration which is
rather moderately-well defined for contact binaries with periods
longer than about 0.5 day (\cite{RD97}).

\subsection{OU Ser}

OU~Ser is the fourth Hipparcos mission discovery in this group
of systems. The light curve shows almost equally
deep minima. With the Hipparcos ephemeride, the system appears
to be an A-type one with a small mass-ratio $q_{sp} = 0.173 \pm 0.017$.
Our spectral classification of the system indicates the spectral
type F9/G0V.

The distinguishing properties of OU~Ser in the Hipparcos database
are its large proper motion and a well measured parallax.
The tangential components of the proper motion are $\mu_{RA} =
-387.5 \pm 0.9$ mas and $\mu{dec} = 2.8 \pm 0.8$ mas. With the
parallax of $p = 17.3 \pm 1.0$ mas this translates into the
spatial components $V_{RA} = -106$ km~s$^{-1}$ and 
$V_{dec} = 1$ km~s$^{-1}$. The mean radial velocity of the
system is $\gamma = -64.08 \pm 0.41$ km~s$^{-1}$. Thus, the
RA and the radial velocity components indicate a high-velocity
star.

The large proper motion of the star had been the reason for
inclusion in the survey by \cite{car94}. They noted also the
broad lines indicating short-period binarity and possibility
of light variations. Their photometric data $V=8.27$,
$(B-V)=0.62$ and $(U-B)=0.08$ agree with our spectral classification,
F9/G0V. The $ubvy$ survey of \cite{ols94} suggests a slightly
larger $(B-V) \simeq 0.66$ on the basis of the $(b-y)=0.411 
\pm 0.003$, hence a spectral type around G1/2V. The difference
in the classification may be due to the apparently low metallicity
of the system as judged by its low index $m_1=0.168 \pm 0.004$
(provided this index is not confused by any chromospheric activity).
the other data of Olsen are $V=8.278 \pm 0.005$ and $c_1 = 0.281
\pm 0.006$.

Assuming $V_{max}=8.25$ and the parallax $p=17.3 \pm 1.0$ 
mas, one obtains $M_V = 4.44 \pm 0.12$. This again agrees very well
with the absolute magnitude derived from the 
$M_V(\log P, B-V)$ calibration of \cite{RD97}:
For $(B-V)=0.62$, $M_V^{cal} = 4.32$, 
while for $(B-V)=0.66$, $M_V^{cal}=4.46$.

With the excellent parallax data and its properties of a high
velocity star, OU~Ser deserves a combined photometric --
spectroscopic solution.

\section{SUMMARY}

The paper brings radial velocity data for the third group of ten 
close binary systems that we observed 
at the David Dunlap Observatory. All, but RZ~Dra (which
had SB1 radial velocity data), have never been
observed spectroscopically; all ten are binaries with 
both components clearly detected so that they can be called SB2. 
All systems, but CN~And and RZ~Dra, which may be very
close semi-detached systems, are contact binaries. 
We describe special features of the individual systems in the
descriptions in Section~\ref{sec2}. We note that again
about half of the system are A-type contact binaries; the likely
reasons why we prefer them over the W-type systems in
our randomly drawn sample are given in the Conclusions to Paper~II.
We also observed as EB2 systems two binaries, CN~And and RZ~Dra;
they are most probably semi-detached binaries.

We do not give the calculated
values of $(M_1+M_2) \sin^3i = 1.0385 \times 10^{-7}\,(K_1+K_2)^3\,
P({\rm day})\,M_\odot$ because in most cases the
inclination angles are either unknown or not trustworthy. However,
one case is very interesting here: The total mass of components
of the system V753~Mon is very large, $M_1+M_2 > 2.93 M_\odot$.
Since {\it too large velocity amplitudes\/} are a rare phenomenon
in the world of contact systems, this system requires special
attention of the observers. The binary is also unique in having its
mass-ratio exceptionally close to unity, $q_{sp}=0.970 \pm 0.009$. 
Two other systems discovered by the
Hipparcos mission are also very important and promise excellent
combined solutions, YY~CrB and OU~Ser. It is important that both
are high-velocity stars and have excellent parallaxes. And, finally,
the recently discovered system HV~Aqr offers an excellent solution
in view of the total, well-defined eclipses and very good radial
velocity data.

\acknowledgements
The authors would like to thank Jim Thomson for help with
observations and to Hilmar Duerbeck for a permission to use
his observations of UX~Eri.

The research has made use of the SIMBAD database, operated at the CDS, 
Strasbourg, France and accessible through the Canadian 
Astronomy Data Centre, which is operated by the Herzberg Institute of 
Astrophysics, National Research Council of Canada.

\clearpage

\noindent
Captions to figures:

\bigskip

\figcaption[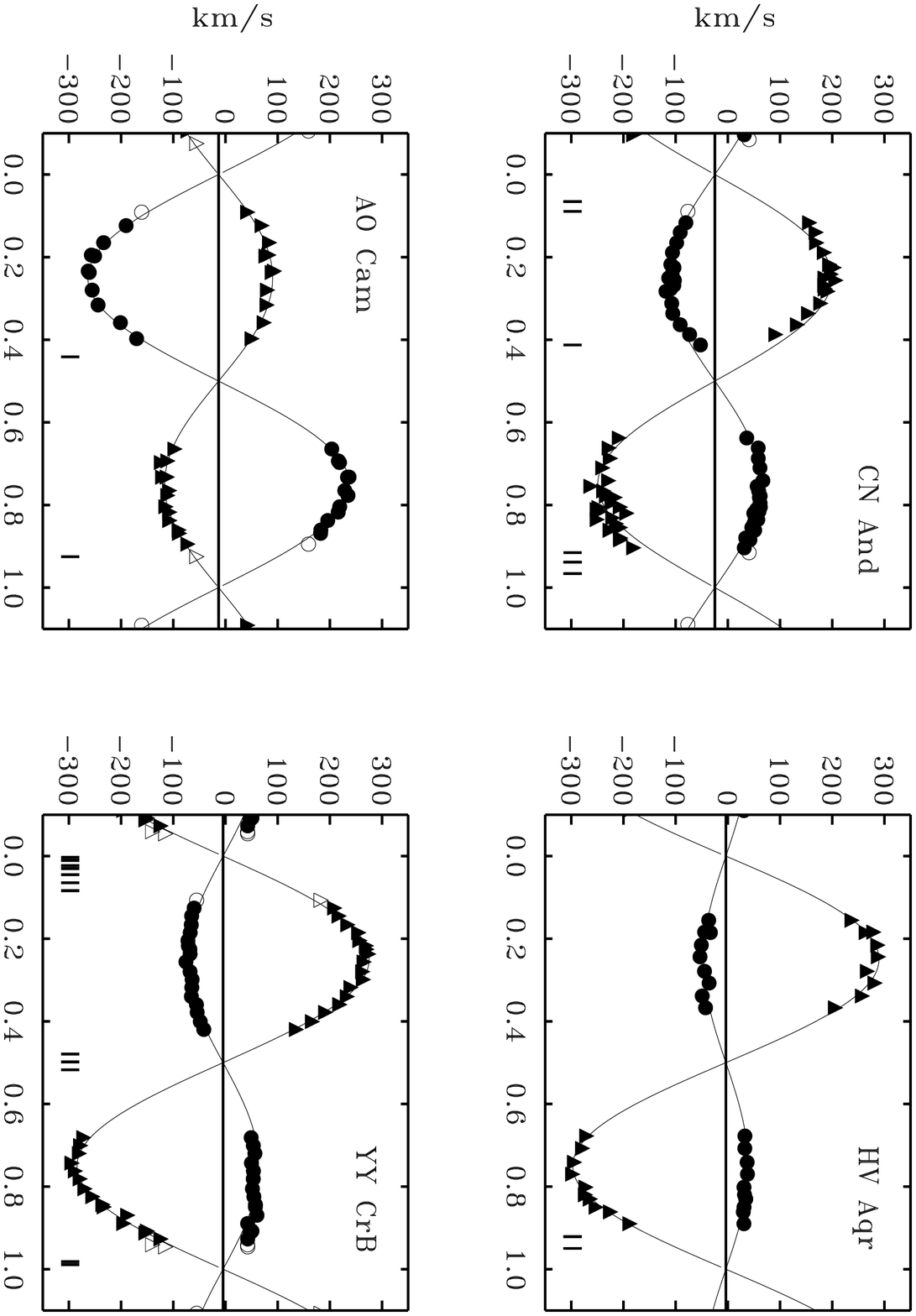] {\label{fig1}
Radial velocities of the systems CN~And, HV~Aqr, AO~Cam and YY~CrB
are plotted in individual panels versus orbital phases. The thin lines 
give the respective circular-orbit (sine-curve) fits to the radial velocities.
Note that AO~Cam is a W-type system while the rest are A-type
systems. Short marks in the lower 
parts of the panels show phases of available 
observations which were not used in the solutions because of the 
blending of lines. Open symbols in this and the next figure
indicate observations given half-weights in the solutions. All
panels have the same vertical scales.
}

\figcaption[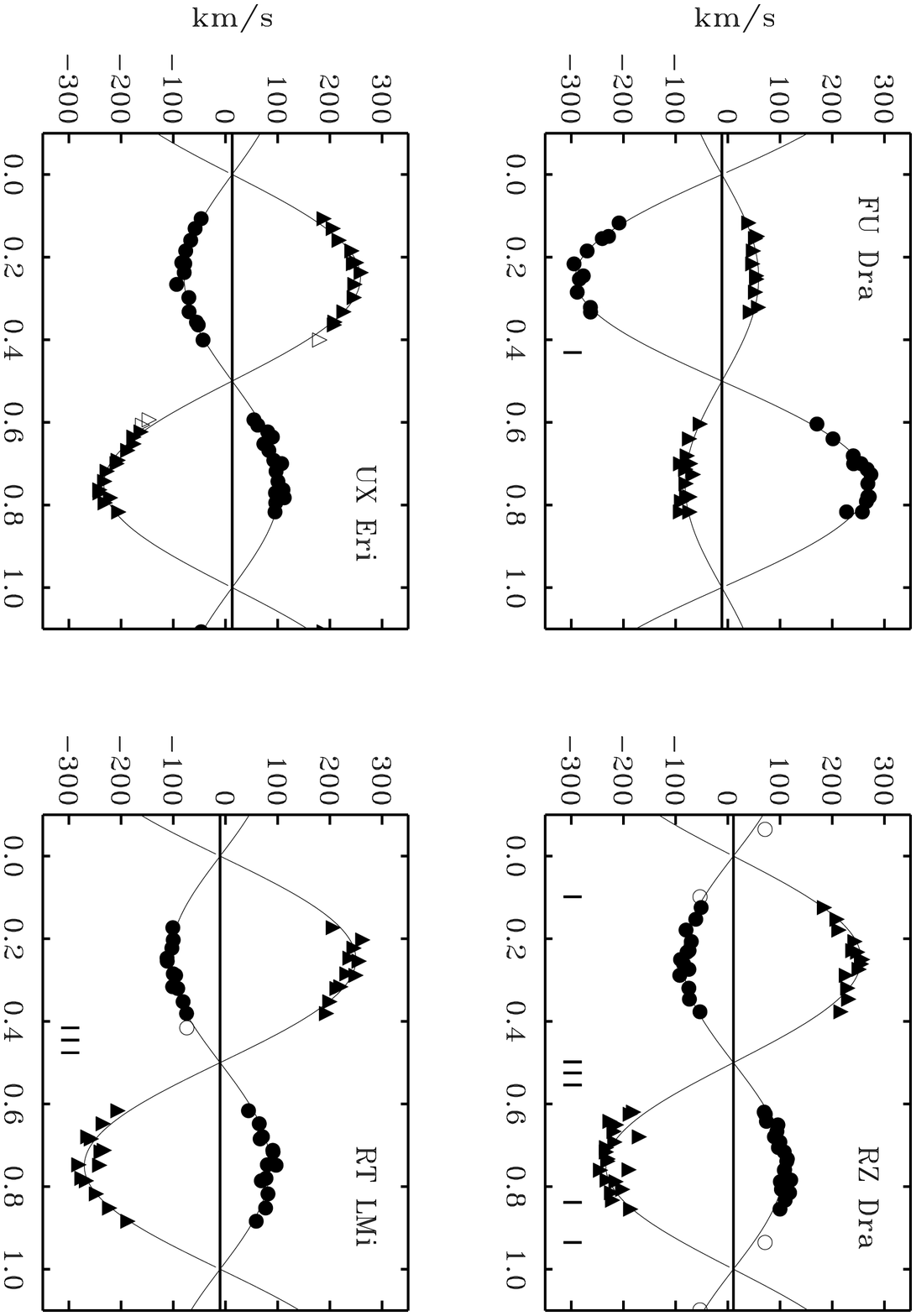] {\label{fig2}
Radial velocities of the systems FU~Dra, RZ~Dra, UX~Eri and RT~LMi. 
}

\figcaption[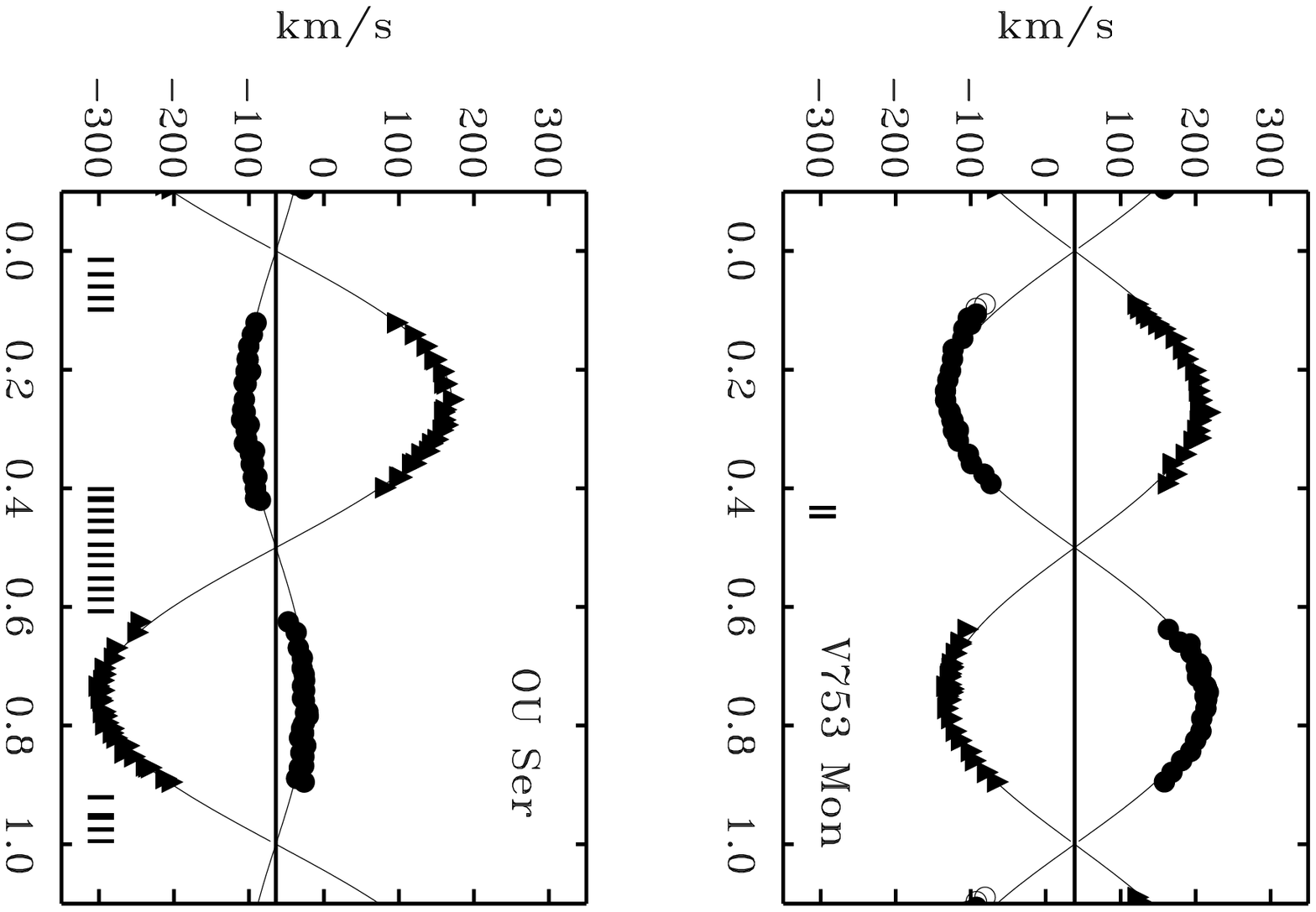] {\label{fig3}
Radial velocities of the systems V753~Mon and OU~Ser. While both of
these recently-discovered systems are very interesting, V753~Mon is
exceptional in having a mass-ratio very close to unity in
showing very large radial-velocity amplitudes.
}

\clearpage

\begin{table}                 
\dummytable \label{tab1}      
\end{table}

\begin{table}                 
\dummytable \label{tab2}      
\end{table}

\end{document}


\pagestyle{empty}

\begin{deluxetable}{lclrrrccc}
\small
\tablecolumns{9}
\tablewidth{0pt}
\tablenum{2}
\tablecaption{Spectroscopic orbital elements of the third
ten of close binary systems}
\tablehead{
   \colhead{Name} &                
   \colhead{Type} &                
   \colhead{Other} &               
   \colhead{$\gamma$} &            
   \colhead{K$_1$} &               
   \colhead{$\epsilon_1$} &        
   \colhead{T$_0$ -- 2,400,000} &  
   \colhead{P (days)} &            
   \colhead{q =} \nl               
   \colhead{Sp.\ type}     &       
   \colhead{}               &      
   \colhead{names} &               
   \colhead{} &                    
   \colhead{K$_2$} &               
   \colhead{$\epsilon_2$} &        
   \colhead{(O--C) and [E]} &      
   \colhead{assumed} &             
   \colhead{$m_2/m_1$}             
}
\startdata
CN~And & A/EB & BD$+39^\circ 59$       & $-24.86$(1.45) & 87.53(1.11)  & 
 6.10  & 50850.2790(22)  & 0.462793    & 0.390(33)    \nl
 ~~~F5V&   &             &             & 224.35(2.84) &
 18.03 & $-0.0137$ [327] &             &              \nl      
HV~Aqr & A    & BD$-3^\circ 5183$      & $-3.77$(1.25)  & 42.41(1.44)  &
 4.83  & 50859.3453(8)   & 0.374460    & 0.145(50)    \nl
~~~F5V &   &             &             & 293.35(2.11) &
 11.46 & \nodata \tablenotemark{a}             &             &              \nl
AO~Cam & W    & BD$+52^\circ 826$       & $-13.17$(0.83) & 250.50(1.05) &
 6.47  & 50960.1131(5)   & 0.329905    & 2.420(11)    \nl
~~~G0V &   &             &             & 103.51(1.28) &
 5.50  & $-0.0121$ [13732] &            & 1/q=0.413    \nl
YY~CrB & A  & HD~141990                & $-4.58$(1.02)  & 68.07(1.54) &
 7.79  & 50955.8711(6)   & 0.376565    & 0.243(23)    \nl
~~~F8V &   & HIP~77598   &             & 279.85(1.55) &
 14.13 & $-0.0053$ [6521] &            &              \nl
FU~Dra & W  &                          & $-11.38$(1.14) & 280.76(2.09) &
 8.68  & 50866.2777(8)   & 0.306718    & 3.989(30)    \nl
~~~F8V &   & HIP~76272   &             & 70.38(1.84)  &
 7.51  & $-0.0086$ [7714] &            & 1/q=0.251    \nl 
RZ~Dra & A/EB &                        & 10.68(1.76)  & 96.85(1.72) &
 7.53  & 51035.4103(28)  & 0.550876    & 0.396(37)    \nl
~~~A6V &   & HIP~90092   &             & 244.47(3.54) &
22.22  & $-0.0165$ [39186] &           &              \nl
UX~Eri & A   & BD$-7^\circ 553$        & 12.79(1.09) & 91.75(1.55) &
 6.81& 50416.5587(14)  & 0.445279    & 0.371(21)   \nl
~~~F9V &   & HIP~14699   &             & 245.76(1.86) &
 10.09 & $+0.0030$ [$-78$] &           &              \nl
RT~LMi & A   &                         & $-10.30$(1.74) & 95.34(1.67) &
 7.28  & 51261.1008(16)  & 0.374918    & 0.366(38)    \nl
~~~F7V:&   &             &             & 260.17(3.59) &
 15.68 & $-0.0016$ [41056] &           &              \nl
V753~Mon & W  & HD~54975               & 38.58(0.83)  & 176.07(0.97) &
 8.84  & 51188.0809(15)  & 0.677049    & 1.031(9)     \nl
~~~A8V &   & HIP~34684   &             & 170.76(0.97) &
 6.80  & $-0.0249$ [3970] &            &  1/q=0.970   \nl
OU~Ser & A& HD~136924                  & $-64.08(0.41)$ & 40.59(0.59) &
 3.80  & 50901.9916(3)  & 0.296764     & 0.173(17)    \nl
~~~F9/G0V& & HIP~75269  &              & 234.24(0.69) &
 5.97  & +0.0026 [8093]  &             &              \nl
\tablenotetext{a}{The two available ephemerides for moments of
primary minima of HV~Aqr give discordant results and do not provide a clear
guidance to which eclipse our $T_0$ applies. It has been
assumed that the system is A-type on the basis of the light curve
data; see the text.}
\tablecomments{The convention of naming the
components is that the subscript 1 designates
the component which is eclipsed at the
deeper minimum and is therefore the hotter one. 
The standard errors of the circular 
solutions in the table are expressed in units of last decimal places 
quoted; they are given in parantheses after each value. 
For example, the last table entry for $q$, 0.173(17) should be 
interpretted as $0.173 \pm 0.017$. 
The average radial velocities ($\gamma$), the 
velocity amplitudes (K$_i$) and the standard unit-weight 
errors of the solutions ($\epsilon$) are expressed 
in km~s$^{-1}$. The calculated moments of primary minima
are given by $T_0$ while their (O--C) deviations (in days) 
have been calculated from the most 
recent available ephemerides, as given in the text,
using the assumed periods and the number of epochs given by [E].}
\enddata
\end{deluxetable}